\documentclass[12pt]{article}

\usepackage{amsmath,amssymb,amsfonts,amsbsy}
\usepackage{cite}
\usepackage{graphicx}
\usepackage{wrapfig}
\usepackage{epsfig}
\usepackage[font=small,labelfont=bf,labelsep=period]{caption}
\begin{document}
\addcontentsline{toc}{subsection}{{The completion of 
single-spin asymmetry measurements at the PROZA setup}\\
{\it V.V. Mochalov}}

\setcounter{section}{0}
\setcounter{subsection}{0}
\setcounter{equation}{0}
\setcounter{figure}{0}
\setcounter{footnote}{0}
\setcounter{table}{0}

\begin{center}
\textbf{THE COMPLETION OF SINGLE-SPIN ASYMMETRY MEASUREMENTS AT 
THE PROZA SETUP}
\vspace{5mm}

\underline{V.V. Mochalov$^{\,1\,\dag}$}, A.N.~Vasiliev$^{\,1}$,
N.A.~Bazhanov$^{\,2}$, N.I.~Belikov$^{\,1}$, A.A.~Belyaev$^{\,3}$,
N.S.~Borisov$^{\,2}$, A.M.~Davidenko$^{\,1}$, A.A.~Derevschikov$^{\,1}$,
V.N.~Grishin$^{\,1}$, A.B.~Lazarev$^{\,2}$, A.A.~Lukhanin$^{\,3}$, 
Yu.A.~Matulenko$^{\,1}$, Yu.M.~Melnik$^{\,1}$, A.P.~Meschanin$^{\,1}$, 
N.G.~Minaev$^{\,1}$, D.A.~Morozov$^{\,1}$, A.B.~Neganov$^{\,2}$, 
L.V.~Nogach$^{\,1}$, S.B.~Nurushev$^{\,1}$, Yu.A.~Plis$^{\,2}$, 
A.F.~Prudkoglyad$^{\,1}$, P.A.~Semenov$^{\,1}$, L.F.~Soloviev$^{\,1}$, 
O.N.~Shchevelev$^{\,2}$, Yu.A.~Usov$^{\,2}$, A.E.~Yakutin$^{\,1}$ \\

\vspace{5mm}

\begin{small}
  (1) \emph{Institute for High Energy Physics, Protvino} \\
  (2) \emph{Joint Institute for Nuclear Research, Dubna} \\
  (3) \emph{Kharkov Physical Technical Institute, Kharkov, Ukraine} \\
  $\dag$ \emph{E-mail: mochalov@ihep.ru}
\end{small}
\end{center} 

\vspace{0.0mm} 

\begin{abstract}
Single spin asymmetry in inclusive $\pi^0$-production 
was measured in the polarized target fragmentation region 
using 50 GeV proton beam. The asymmetry is in agreement 
with asymmetry measurements in the polarized beam 
fragmentation region carried out at higher energies. 
The measurement completed 30-years history of polarized 
measurements at the PROZA setup.
\end{abstract}

\vspace{2.2mm} 

\subsubsection*{Introduction}
This report is sad in some sense, because it concludes 
the 30 year's history of PROZA experiment. Nevertheless the
new experimental program is planning at IHEP (see S.~Nurushev's
talk \cite{nur_spin09}). 
We present the recent results in inclusive $\pi^0$ production 
as well as the highlights of the previous polarization 
study at PROZA. 

\vspace{-2mm} 
 
\subsubsection*{Asymmetry in the unpolarized beam 
fragmentation region}
\vspace{-2mm}

\begin{figure}[b!]
  \centering
    \includegraphics[width=0.7\textwidth]{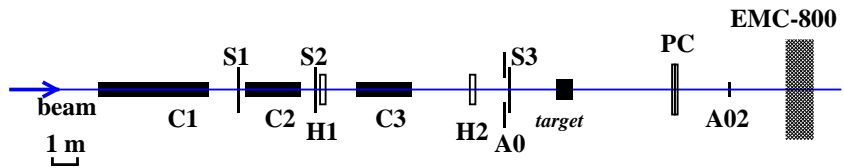} 
  \caption{%
    \small The Experimental Setup PROZA, $S1$-$S3$-- trigger 
scintillator counters, $A0,A02$--beam anti-coincidence counters,
$H1$-$H2$-- beam hodoscopes, $PC$--Proportional chamber, 
$EMC$-$800$ -- Electromagnetic calorimeter}
  \label{fig1_forw_setup}
\end{figure}

\begin{wrapfigure}[12]{R}{70mm}
  \centering 
  \vspace{-2mm} 
  \includegraphics[width=70mm]{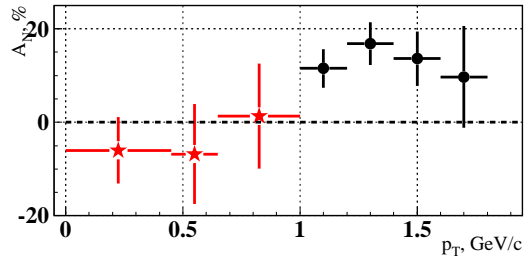}
  \caption{
\small 
$A_N$ in the reaction 
$\pi^{-}+d_{\uparrow} \rightarrow\pi^{0}+X$ in the beam 
fragmentation region. Circles -- currents measurements, stars --
previous data \cite{forwd,forwp}}
  \label{fig:asym_forw}
\end{wrapfigure}

Single (left-right) spin asymmetry was measured in the reaction 
$\pi^{-}+d_{\uparrow} \rightarrow\pi^{0}+X$ in the beam 
fragmentation region. The experimental Setup is presented
in Fig.~\ref{fig1_forw_setup}. $\gamma$-quanta were measured 
using a lead-glass electromagnetic calorimeter 
placed at 8~m downstream the target.

The asymmetry $A_N^{meas} (\phi)$ was calculated 
for each angle $\phi$ as the difference of the 
normalized numbers of pions $n_{\downarrow \uparrow}$ 
(equivalent to the differential cross-section) 
for opposite signs of target polarization:

$A_N^{meas} (\phi) =\frac{D}{P_{targ}}\cdot A_N^{raw} =
\frac{D}{P_{targ}}\cdot 
\frac{n_{\uparrow}-n_{\downarrow}}{n_{\uparrow}+n_{\downarrow}}$

An average polarization value ($P_{targ}$) of fully deuterized 
propane-diol target was 35\%, dilution factor D=2.5 to 5 
decreasing with $x_F$ increases. The procedure to measure
carefully the dilution factor is described in detail 
elsewhere\cite{forwd}. The asymmetry sign was 
selected to be consistent with all polarized 
beam experiments.

Final asymmetry $A_N$ was calculated by fitting the measured 
asymmetry by linear function $A_N^{meas} (\phi) = A_0 + A_N \cdot cos (\phi)$.

The last procedure allowed to eliminate  
systematic errors caused by beam monitor instability. 
Asymmetry was measured in the range of $0.6< x_f <1.0$ and 
$1.0<p_T<2.0$~GeV/c. 

The results are presented in Fig.~\ref{fig:asym_forw} and in 
Table~\ref{tab:pt}.

\begin{table}[h]
\begin{center}
\caption{
\small
$A_N$ in the reaction $\pi^{-}+d_{\uparrow} \rightarrow\pi^{0}+X$
}
\begin{tabular}{|l||c|c|c|c||c|}
\hline
\hline
$x_F$ / $p_T$, GeV/c & 1.0-1.2 & 1.2-1.4 & 1.4-1.6 & 1.8-2.0 & 1.0-2.0 \\
\hline 
\hline
0.6-0.7 & $ -8.6 \pm 8.0$ & $ 2 \pm 10.0$ & $ -23.0 \pm 16.0$ &
$ 2.0 \pm 25.0$ & {\boldmath $ -6.5 \pm 5.7 $} \\
\hline
0.7-0.8  & $ 17.0 \pm 8.0$ & $ 30.0 \pm 9.5$ & $ 24.0 \pm 11.0$ &
$ 6.0 \pm 16.0$ & {\boldmath $ 21.1 \pm 5.1 $} \\
\hline
0.8-0.9  & $ 7.0 \pm 6.6$ & $ 19.0 \pm 8.0$ & $ 8.9 \pm 8.7$ &
$ 13.0 \pm 15.0$ &  {\boldmath $ 11.3 \pm 4.2 $} \\
\hline
0.9-1.0  & $ 12.4 \pm 6.9$ & $ 8.0 \pm 7.0$ & $ 10.7 \pm 10.8$ &  
& {\boldmath $ 10.3 \pm 4.5 $} \\
\hline
\hline
 0.7-1.0  & $ 11.5 \pm 4.1$ & $ 16.8 \pm 4.6$ & $ 13.6 \pm 5.8$ &
$ 9.7 \pm 10.9$ & {\boldmath $ 13.6 \pm 2.6 $} \\
\hline
\hline
\end{tabular} 
\label{tab:pt}
\end{center}
\end{table}

Surprisingly asymmetry is notable near the edge of phase
space: 

$A_N=(13.6\pm 2.6(stat) \pm 2.0(syst))\%$ 
at $p_T>1$ (GeV/c) and $0.7< x_F<1.0$. 

Non-zero asymmetry in the unpolarized beam fragmentation 
region can be explained in some models\cite{spin86,lapidus}.
The asymmetry value in unpolarized $\pi^-$-beam 
fragmentation region close to the one found in the 
same reaction in the polarized target 
fragmentation region\cite{piback}, and also 
to the neutron polarization measured in the reaction 
$\pi^{-}+p_{\uparrow} \rightarrow\pi^{0}+n$ at the same 
transferred momentum \cite{exclus_pi}.

Asymmetry was also measured in the reaction 
$K^{-}+d_{\uparrow} \rightarrow\pi^{0}+X$ at $x_F>0.7$. 
$A_N=(-0.4 \pm 6.1)\%$ at $p_T (GeV/c) \leq 1.2$ and
$A_N=(11\pm 6.1)\%$ at $p_T>1.2$ (GeV/c).

\begin{figure}[t!]
  \centering
  \begin{tabular}{cc}
    \includegraphics[width=75mm]{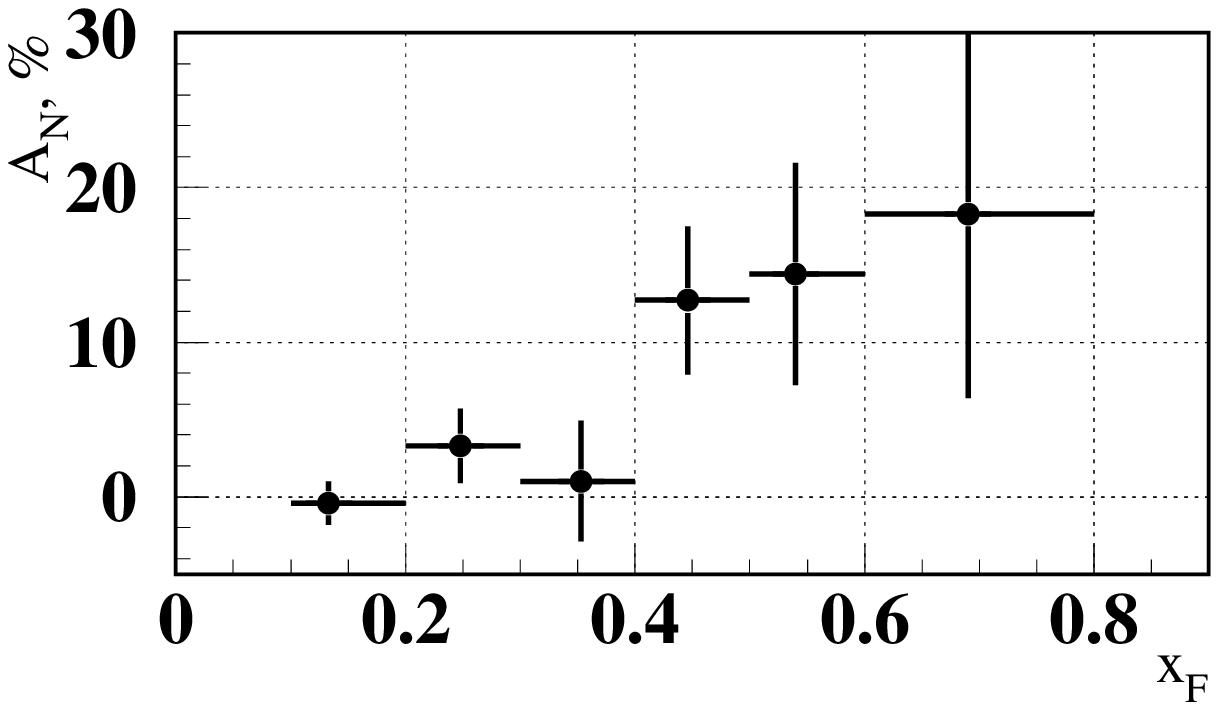} &
    \includegraphics[width=75mm]{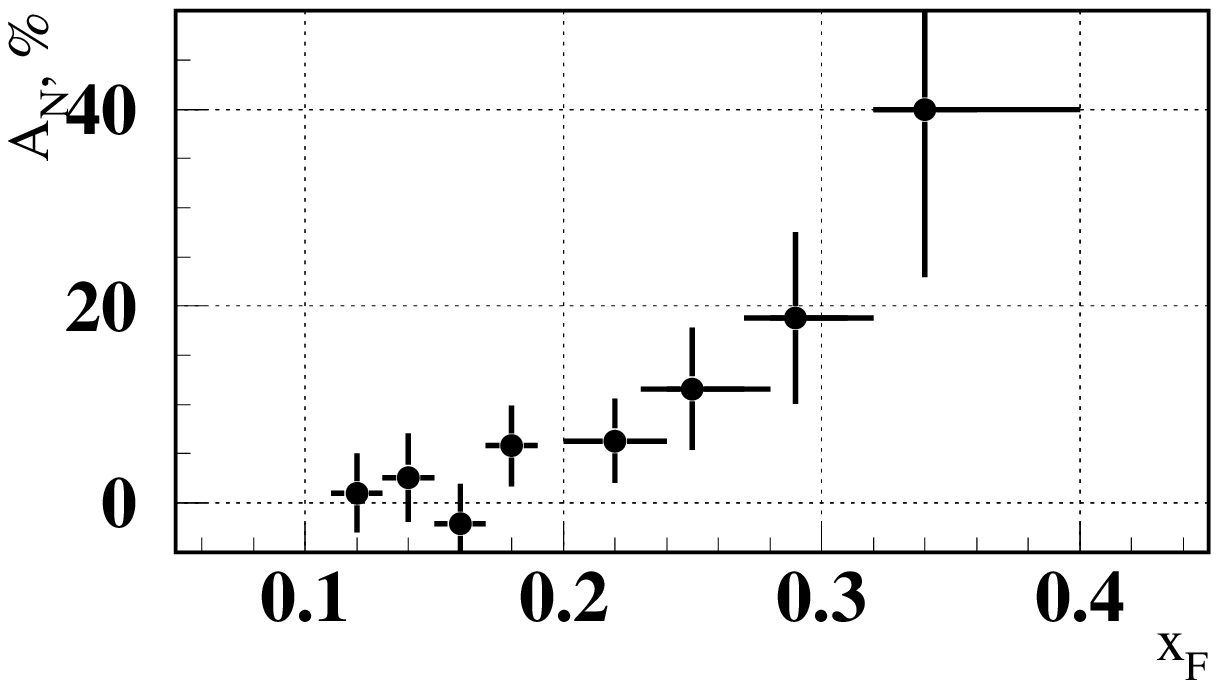} \\
    \textbf{(a)} & \textbf{(b)}
  \end{tabular}
  \caption{%
\small
$A_N$ in the polarized target fragmentation region:
    \textbf{(a)} in the reaction 
$\pi^{-}p_{\uparrow} \rightarrow\pi^{0}X$ at 40 GeV\cite{piback};
    \textbf{(b)} in the reaction 
$pp_{\uparrow} \rightarrow\pi^{0}X$ at 70 GeV\cite{pback}. 
Results are presented in the polarized proton  
fragmentation region to be consistent with the polarized beam data.}
  \label{fig:asym_back}
\end{figure}

\subsubsection*{Asymmetry in the polarized target 
fragmentation region}

\begin{figure}[b!]
  \centering
  \begin{tabular}{cc}
    \includegraphics[width=60mm]{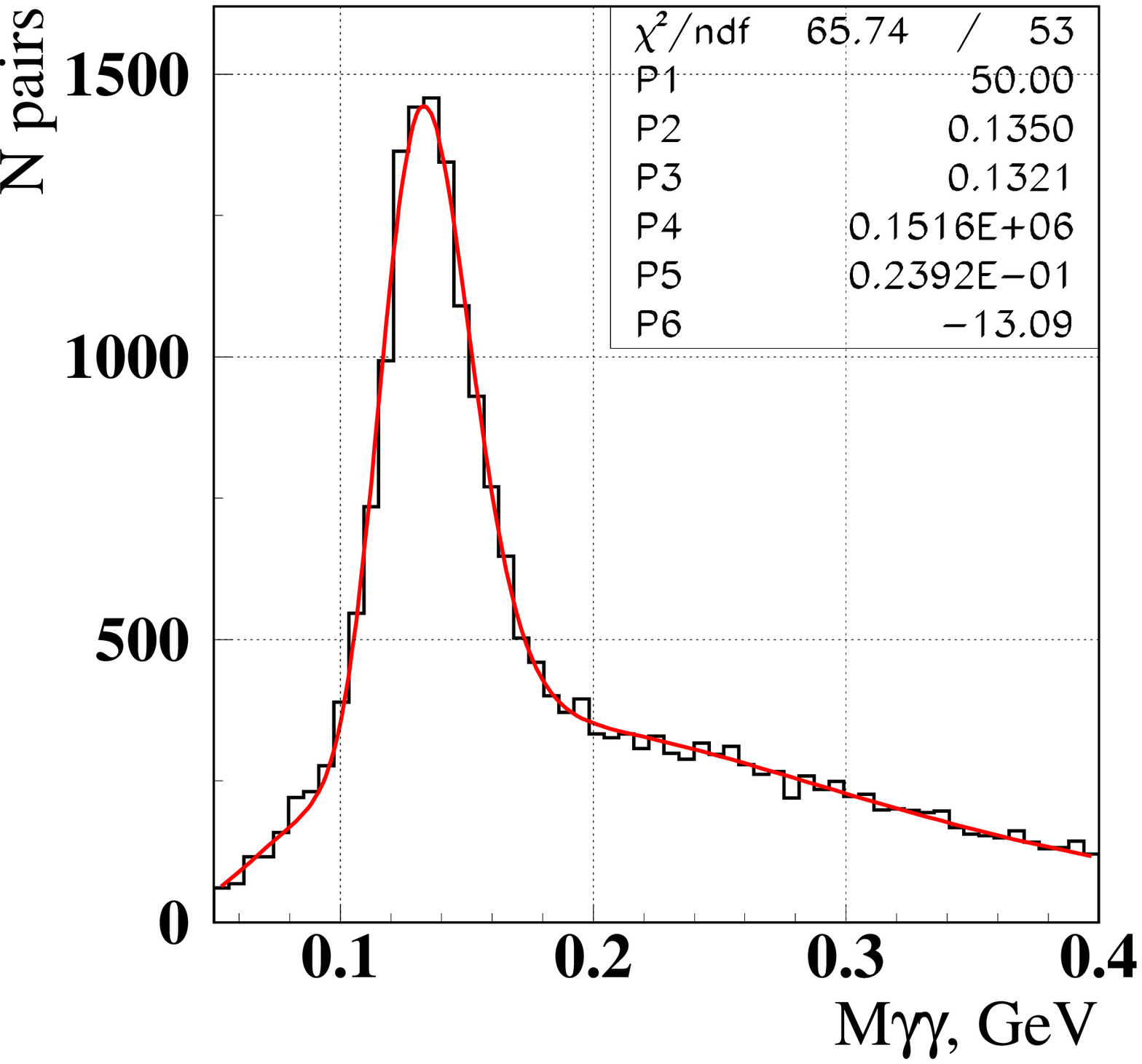} &
    \includegraphics[width=90mm]{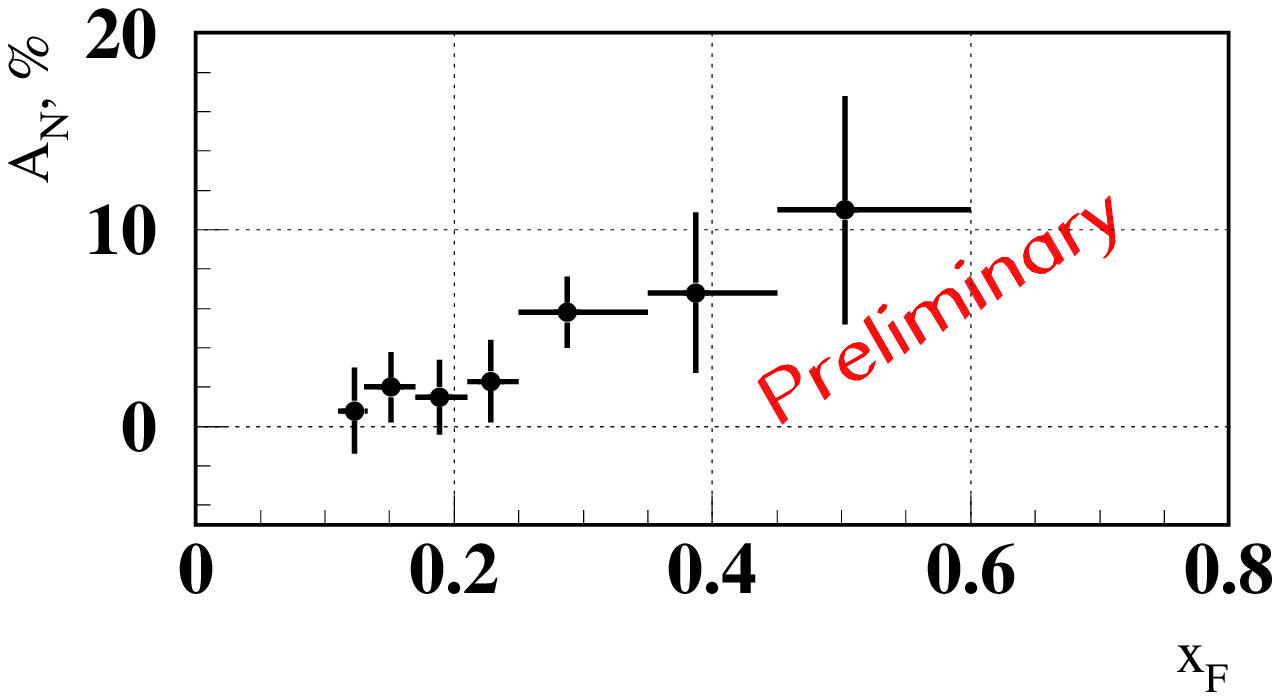} \\
    \textbf{(a)} & \textbf{(b)}
  \end{tabular}
  \caption{%
\small
    \textbf{(a)} $\gamma \gamma$-mass spectrum;
    \textbf{(b)} $A_N$ in the reaction 
$p+p_{\uparrow} \rightarrow\pi^{0}+X$ at 50 GeV. 
}
  \label{fig:asym_new}
\end{figure}

$A_N$ in the polarized target fragmentation region 
was measured earlier at PROZA at 40~GeV \cite{piback} 
and 70~GeV\cite{pback} (see Fig.~\ref{fig:asym_back}).
We present an asymmetry measurement in the 
reaction $p+p_{\uparrow} \rightarrow\pi^{0}+X$ 
in the polarized target fragmentation region at 50 GeV. 
Two sets of data (2005 and 2007) are being used for analysis. 
The experiment was carried out at the upgraded PROZA-2M setup.
The electromagnetic calorimeter was placed at 2.3~m 
downstream the target at $30^{\circ}$ respect to the 
beam direction. The geometry was
selected to detect neutral pions in the backward hemisphere. 
A special trigger on transverse momentum $p_T$ allowed to enrich
data with negative values of $x_F$. 
A special algorithm was developed to reconstruct $\gamma$'s
which hit the detector at large angles 
(up to 20 degrees)\cite{solo}. The $\pi^0$ mass spectrum
is presented in Fig.~\ref{fig:asym_new}a. 

A single-arm experimental setup was used. A special procedure 
was used to eliminate systematic errors \cite{piback}.
The asymmetry was measured at $-0.6<x_F<-0.2$. The result is 
presented in Fig.~\ref{fig:asym_new}b and in Table.~\ref{tab:xf}. 
$x_F$ and asymmetry values are inverted to be consisted 
with the existing polarized beam data.

\begin{table}[h]
\begin{center}
\caption{
\small
$A_N$ in the reaction $p+p_{\uparrow} \rightarrow\pi^{0}+X$ at 50 GeV}
\begin{tabular}{|l||c|c|c|c|c|c|c|}
\hline
$x_F$ & 0.11-0.13 & 0.13-0.17 & 0.17-0.21 & 0.21-0.25
 & 0.25-0.35 & 0.35-0.45 & 0.45-0.60 \\
\hline 
$A_N$,\% & $ 0.8 \pm 2.2$ & $2.0 \pm 1.8$ & $ 1.5 \pm 1.9$ &
$ 2.3 \pm 2.1$ & $5.8 \pm 1.8$ & $6.8 \pm 4.1$ & $11.0 \pm 5.8$\\
\hline
\end{tabular} 
\label{tab:xf}
\end{center}
\end{table} 

The asymmetry measured at $-0.6<x_F<-0.25$ ($6.2\pm 1.5\%$) is 
in a very good agreement with the previous PROZA data at 40 GeV
($6.9\pm 2.8\%$)\cite{piback}, the E704 
data ($6.3\pm 0.7\%$)\cite{e704} and with the STAR data\cite{star}.
Single-spin asymmetry does not depend on energy in a 
very wide range. Intermediate energies give us the possibility
to measure asymmetry of a variety types of particles with excellent accuracy.

\subsubsection*{Highlights of the previous PROZA results}

\begin{figure}[h!]
  \centering
  \begin{tabular}{cc}
    \includegraphics[width=70mm,height=40mm]{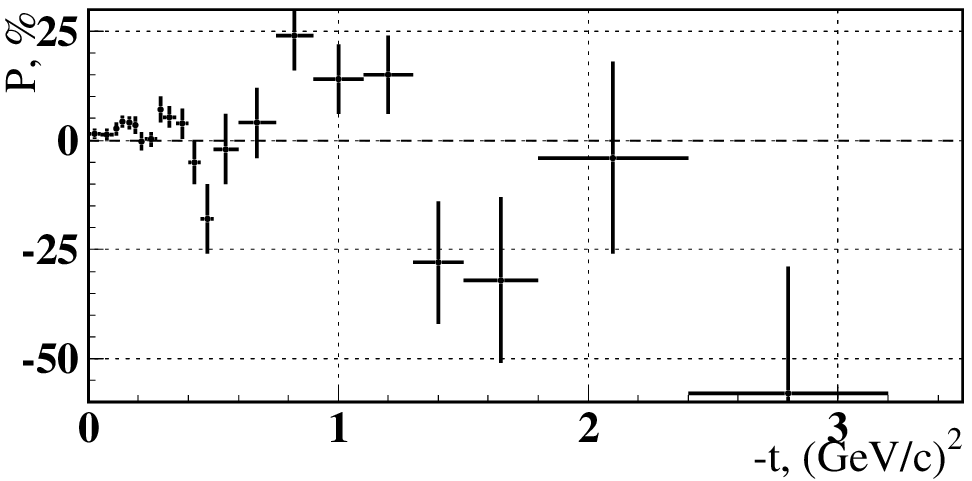} &
    \includegraphics[width=80mm,height=40mm]{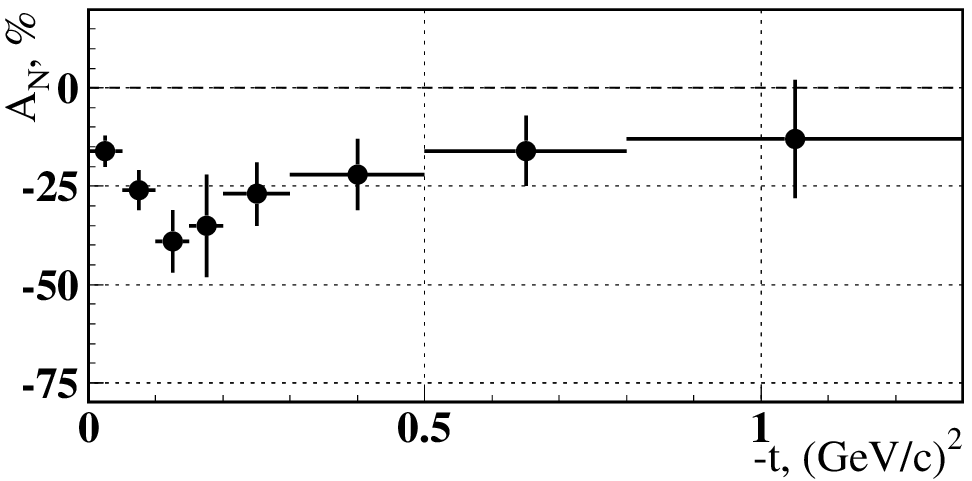} \\
    \textbf{(a)} & \textbf{(b)}
  \end{tabular}
  \caption{%
\small
    \textbf{(a)} Polarization in the reaction 
$\pi^-+p_{\uparrow} \rightarrow\pi^{0}+n$\cite{excl1}. 
    \textbf{(b)} $A_N$ in the reaction
$\pi^-+p_{\uparrow} \rightarrow f_2 (1270)+n$ at 70 GeV\cite{excl2}. 
}
  \label{fig:exclus}
\end{figure}

Asymmetry was measured in different exclusive 
charge-exchange reactions: 
$\pi^-p_{\uparrow}\rightarrow \pi^0(\eta ,\eta \prime (958),
w(783),f_2 (1270))n$ at 40 GeV\cite{excl1}-\cite{excl4}. 
These results are presented in another talk\cite{nur_spin09}.
Only two out of several ones results are presented 
in Fig.~\ref{fig:exclus}

\begin{figure}[b!]
  \centering
  \begin{tabular}{cc}
    \includegraphics[width=70mm,height=40mm]{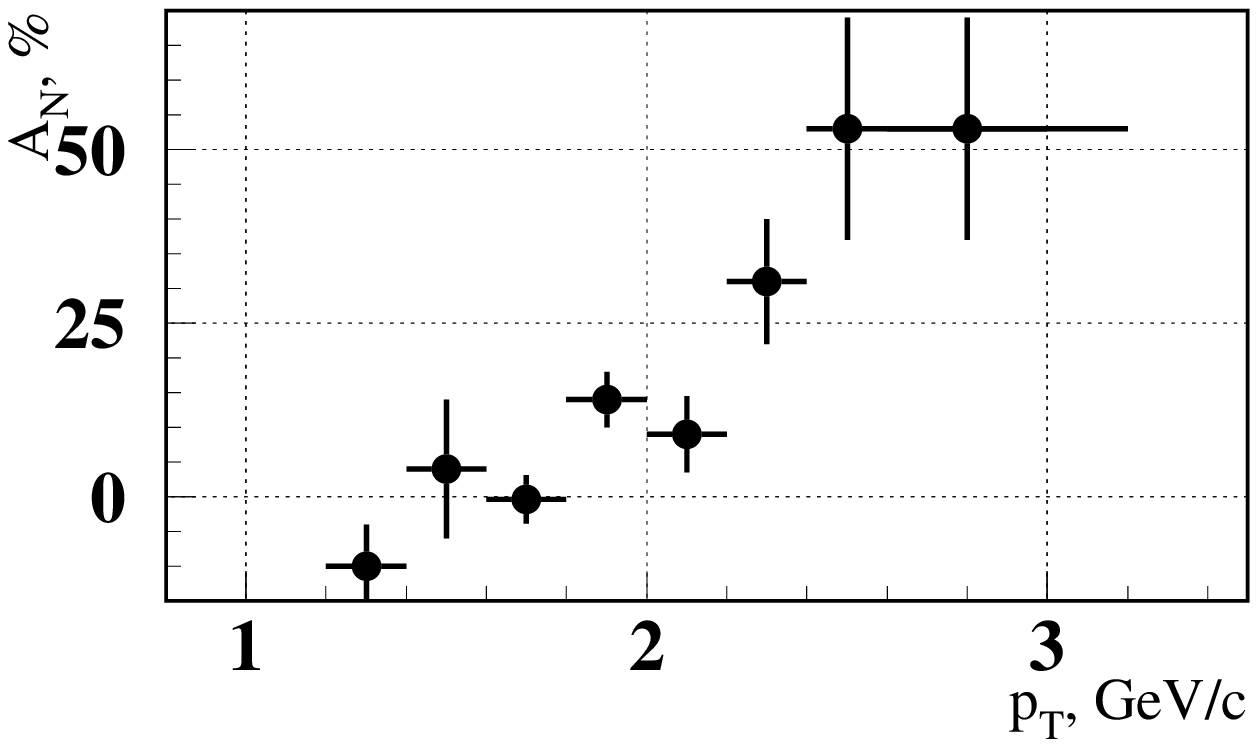} &
    \includegraphics[width=80mm,height=40mm]{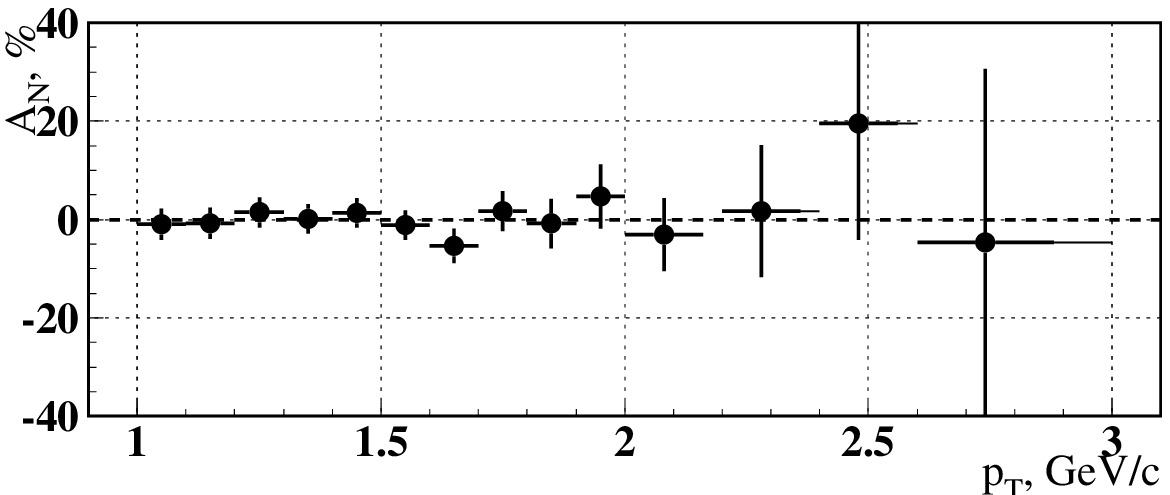} \\
    \textbf{(a)} & \textbf{(b)}
  \end{tabular}
  \caption{%
\small
$A_N$ in the central region in the reactions:
    \textbf{(a)} 
$\pi^-+N(p,d)_{\uparrow} \rightarrow\pi^{0}+X$ at 40 GeV\cite{incl1}. 
    \textbf{(b)} 
$p+p_{\uparrow} \rightarrow\pi^{0}+X$ at 70 GeV\cite{incl2}.
Asymmetry sign is inverted to be consistent with all polarized beam
data. 
}
  \label{fig:asym_cent}
\end{figure}

Let me remind here the most interesting features of the
observed polarization (asymmetry) behavior:
\begin{itemize}
\vspace{-3mm}
\item polarization has a minimum when differential 
cross-section changes it's slope;
\vspace{-3mm}
\item there are oscillations in polarization behavior;
\vspace{-3mm}
\item there is an indication that asymmetry is bigger 
for heavier particles.
\end{itemize}

Lessons from these data will be discussed later.
All these asymmetries were measured 
more than 25 years ago, nevertheless there are no 
theoretical models describing all the data together.

PROZA experiment was one of the first to measure 
single spin asymmetry in inclusive production. 
$A_N$ in inclusive $\pi^0$-production in the central region
is presented in Fig.~\ref{fig:asym_cent}.

Unexpectedly large single-spin asymmetry in $\pi^0$-production
at $\pi^-$-beam in the central region \cite{incl1} pointed 
out that polarization effects are beam quark flavour dependent, since
$A_N$ in $pp_{\uparrow}$ interactions is zero at the same energy.

\subsubsection*{Answers and {\it question} instead of Conclusion}
PROZA experiment found many interesting effects both in 
exclusive and inclusive reactions. Nevertheless 
even {\it more questions have to be discussed}.

Let's first summarize what we know and what we {\it can not
explain} in exclusive reactions.

\begin{itemize}
\item A significant polarization (asymmetry) was found in the all 
exclusive reactions\cite{excl1}-\cite{excl4}. 
{\it Does the asymmetry magnitude increase with meson mass?}
\item There is an indication on asymmetry oscillations.
{\it Is it real effect for all particles? 
Better accuracy is required.}
\item
Polarization changes it's sign in the dip region 
on the $\pi^-+p_{\uparrow} \rightarrow\pi^{0}+n$ 
differential cross-section.
{\it Is it valid for other reactions? What is the 
theoretical explanation of this effect?}
\item Simple Regge model can not describe 
polarization. Modification was required. One of the possible 
solution is Odderon pole in addition to $\rho$-pole.
{\it There is no predictions for the most of the reactions} 
except $\pi^-+p_{\uparrow} \rightarrow\pi^{0}+n$ 
(see \cite{golosk} for example).
{\it Another interesting prediction is that 
$P(\pi^0)+2P(\eta)=P(\eta\prime)$. It is very interesting 
also to measure asymmetry in $a_0(980)$ 
production \cite{achasov}).}
\end{itemize} 

Similar complicated situation is for inclusive reactions.
\begin{itemize}
\item Asymmetry mainly does not depend on 
energy (see also E704, BNL, RHIC data)
{\it We have very good possibility to measure 
asymmetry in different channels at intermediate 
energies with good accuracy at the SPASCHARM experiment.}
\item A significant asymmetry was found for $u-$ and $d-$ 
quark particles. The asymmetry is quark flavor dependent 
(at least for pion and proton beams). The asymmetry in the
$\eta$-production is bigger than in the $\pi^0$ production 
(see also STAR data).
{\it What is the asymmetry for $ss-bar$ and heavier 
states ($\phi$ and others)?}.
\item Asymmetry increases with $p_T$ at the central region 
in the reaction $\pi^-+p_{\uparrow} \rightarrow\pi^{0}+X$
{\it Most of the models can not predict non-zero 
asymmetry in the central region and describe $p_T$ behavior.}
\item A threshold effect and a scaling was observed.
Asymmetry in the non-polarized beam and in the polarized 
target (beam) regions close to the edge of phase 
space are equal each other in the reaction 
$\pi^-+p_{\uparrow} \rightarrow\pi^{0}+X$
{\it It is very important to measure asymmetry 
in a wide kinematic region in different channels 
to discriminate between different models.}
\end{itemize}

We may conclude that we have found a lot of interesting 
spin effects. Nevertheless we all have desires, 
possibilities and duties trying to find much more 
inviting and unpredictable.

\subsubsection*{Acknowledgement}

The work was supported by State Atomic Energy 
Corporation Rosatom with partial support by State 
Agency for Science and Innovation grant 
N 02.740.11.0243 and RFBR grants 08-02-90455 
and 09-02-00198.

\end{document}